\documentclass{article}
\usepackage[a4paper, margin=1.25in]{geometry}
\usepackage[utf8]{inputenc}
\usepackage[style=authoryear-comp,maxcitenames=2,uniquelist=false,backend=biber,natbib,maxbibnames=99,giveninits=true,dashed=false,uniquename=init]{biblatex}
\DeclareNameAlias{sortname}{family-given} % ALways write names like lastname, firstname
\usepackage[dvipsnames]{xcolor}
\usepackage{soul}
\usepackage{siunitx}
\usepackage{graphicx}
\usepackage{amsmath}
\usepackage{bm}
\usepackage{url}
\usepackage{authblk}
\usepackage{hyperref}
\usepackage{xcolor}
\hypersetup{
    colorlinks,
    linkcolor={red!50!black},
    citecolor={blue!50!black},
    urlcolor={blue!80!black}
}

\DefineBibliographyStrings{english}{
  mathesis = {MSc thesis},
}

\title{Joint 2D to 3D image registration workflow for comparing multiple slice photographs and CT scans of apple fruit with internal disorders}
\author[1,*]{Dirk Elias Schut}
\author[2]{Rachael Maree Wood}
\author[3]{Anna Katharina Trull}
\author[4]{Rob Schouten}
\author[1,5]{Robert van Liere}
\author[1,6]{Tristan van Leeuwen}
\author[1,7]{Kees Joost Batenburg}

\affil[1]{Computational Imaging Group, Centrum Wiskunde en Informatica (CWI), Science Park 123, 1098 XG Amsterdam, The Netherlands}
\affil[2]{Horticulture and Product Physiology, Wageningen University and Research, Droevendaalsesteeg 1, 6708 PB Wageningen, The Netherlands}
\affil[3]{Greefa Machinebouw B.V., Langstraat 12, 4196 JB Tricht, The Netherlands }
\affil[4]{Wageningen Food and Biobased Research, Bornse Weilanden 9, 6708 WG Wageningen, The Netherlands}
\affil[5]{Visualization Group, Eindhoven University of Technology, PO Box 513, 5600 MB Eindhoven, The Netherlands}
\affil[6]{Mathematisch Instituut, Utrecht University, Budapestlaan 6, 3584 CD Utrecht, The Netherlands}
\affil[7]{Leiden Institute of Advanced Computer Science (LIACS), Leiden University, Niels Bohrweg 1, 2333 CA Leiden, The Netherlands}
\affil[*]{Corresponding author. E-mail: dirk.schut@cwi.nl}

\date{}

\begin{document}

\maketitle
\clearpage
%\linenumbers

\section*{Abstract}
%A concise and factual abstract is required. The abstract should state briefly the purpose of the research, the principal results and major conclusions. An abstract is often presented separately from the article, so it must be able to stand alone. For this reason, References should be avoided, but if essential, then cite the author(s) and year(s). Also, non-standard or uncommon abbreviations should be avoided, but if essential they must be defined at their first mention in the abstract itself.

A large percentage of apples are affected by internal disorders after long-term storage, which makes them unacceptable in the supply chain. CT imaging is a promising technique for in-line detection of these disorders. Therefore, it is crucial to understand how different disorders affect the image features that can be observed in CT scans. This paper presents a workflow for creating datasets of image pairs of photographs of apple slices and their corresponding CT slices. By having CT and photographic images of the same part of the apple, the complementary information in both images can be used to study the processes underlying internal disorders and how internal disorders can be measured in CT images. The workflow includes data acquisition, image segmentation, image registration, and validation methods. The image registration method aligns all available slices of an apple within a single optimization problem, assuming that the slices are parallel. This method outperformed optimizing the alignment separately for each slice. The workflow was applied to create a dataset of 1347 slice photographs and their corresponding CT slices. The dataset was acquired from 107 ‘Kanzi’ apples that had been stored in controlled atmosphere (CA) storage for 8 months. In this dataset, the distance between annotations in the slice photograph and the matching CT slice was, on average, $1.47 \pm 0.40\;\unit{\milli\meter}$. Our workflow allows collecting large datasets of accurately aligned photo-CT image pairs, which can help distinguish internal disorders with a similar appearance on CT. With slight modifications, a similar workflow can be applied to other fruits or MRI instead of CT scans.

\textbf{Keywords:} Image registration, internal browning, non-destructive testing (NDT), transformation model, automatic differentiation, deep learning

\section{Introduction}
%State the objectives of the work and provide an adequate background, avoiding a detailed literature survey or a summary of the results.

Apples (Malus × domestica Borkh.) are often stored long-term to supply markets with a high-quality product throughout the year \citep{wood2022apple}. Long-term storage of apples is possible by utilizing cold storage under a controlled atmosphere (CA) where oxygen is lowered and carbon dioxide is increased \citep{neuwald2021impact}. However, these conditions involve a loss of fruit quality, allowing physiological disorders and decay to develop. Postharvest losses caused by physiological disorders and decay can range between 18-27 \% under commercial CA conditions \citep{argenta2021characterization}. Therefore, detecting and removing affected fruit from the supply chain is crucial for reducing food wastage. However, disorders such as internal browning(IB) do not display any external disorders making in-line detection challenging. Furthermore, different patterns of IB occur in different regions of fruit tissue \citep{sidhu2023internal}. IB can cause ~34 \% of postharvest losses in sensitive cultivars even when stored under optimal CA conditions \citep{wood2022seasonal}. Accurately detecting internal disorders and defects is critical for maintaining consumer confidence.

X-ray-based imaging techniques are promising for the in-line detection of internal disorders. There are two main X-ray-based imaging modalities: radiography, which uses a single 2D X-ray image (radiograph), and computed tomography (CT), which combines the data of many radiographs of the same object to create a 3D representation of the object. CT scanning offers 3D information and higher contrast, but radiographs can be acquired much faster. While both modalities are promising, radiography currently has more potential for practical applications in in-line quality control of produce, because it has a lower acquisition time \citep{kotwaliwale2014x}. Nevertheless, the acquisition time of CT scanning can be lowered by optimizing the scanning setup for in-line use \citep{morton2009ultrafast, de2016line, schut2022top}. Moreover, radiographs can be accurately simulated from CT scans \citep{van2016fast}, and if labels or segmentation masks are available on the CT scan, these can also be transferred to the simulated radiograph. \citet{zeegers2022tomographic} developed a workflow for foreign object detection on radiographs where a few CT scans with segmentation masks of the foreign objects were used to simulate many radiographs with segmentation masks to train a neural network. This method was extended and applied to predict the ripeness of avocados by \citet{andriiashen2023ct}.

CT scanning of apples has been used for studying the processes underlying bruising \citep{DIELS201724}, internal browning \citep{HERREMANS2013114, CHIGWAYA2021111464}, watercore \citep{HERREMANS201442} and bitterpit \citep{si2016computed, jarolmasjed2016postharvest}. However, several conditions can look similar in CT scans, so destructive methods may still be needed to classify specific disorders. Watercore in apples appears as high-intensity regions in CT images due to the flooding of pores in the water-soaked regions of affected fruit \citep{HERREMANS201442}. High-intensity regions are also reported to be associated with IB \citep{chigwaya2021use}. In some markets, apples with watercore are sold at a premium due to the sweet juicy taste \citep{HERREMANS201442, mink1973apple}, whereas IB is associated with off-flavor and is an unwanted trait \citep{hatoum2014effects}. Furthermore, watercore can dissipate during storage in mildly affected fruit \citep{upchurch1994effects}, while IB will not. Similarly, bruises and bitter pit can appear as small low-intensity regions close to the fruit peel in CT scans \citep{DIELS201724, jarolmasjed2016postharvest}. Bitter pit progressively worsens with CA storage \citep{jarolmasjed2016postharvest}, while fruit with mechanical damage or bruising can still remain healthy during long-term storage. Therefore, fruits must still be cut and visually examined for further development of detection algorithms for specific disorders.

\begin{figure}[htb]
    \centering
    \includegraphics[width=0.6\textwidth]{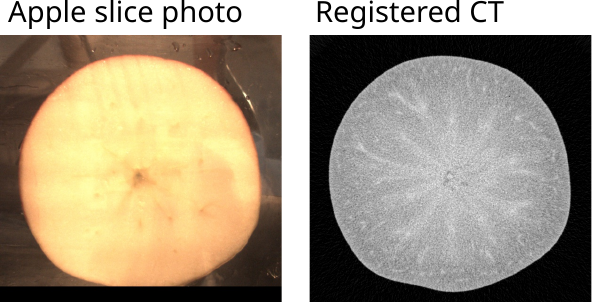}
    \caption{Registered image pair of a slice photograph and a CT slice generated using our method.}
    \label{fig:photo_ct_comparison}
\end{figure}

When fruit are CT or MRI scanned for research, they are often also cut and photographed. It is helpful to present the photographic images alongside CT or MRI images, which requires aligning the photographs to the CT or MRI scan. Photographs have been aligned manually to CT or MRI scans for investigating disorders in apples \citep{CHIGWAYA2021111464, HERREMANS201442, VANDAEL2019218} and pears \citep{VANDELOOVERBOSCH2021114925, azadbakht2019relation, VANDELOOVERBOSCH2020107170, VANDAEL201733}. However, there are significant downsides to manually aligning the photographs. Firstly, manually aligning photographs is tedious and time-consuming, which limits the dataset size. It is often unknown whether an apple has an internal defect before CT scanning, and there is natural variation in how internal defects affect the CT measurements. Therefore, acquiring large datasets is valuable because it increases the chance of capturing the variation. Moreover, having large and varied datasets helps with training machine learning methods. Secondly, manual alignment is challenging in slices without features that are clearly visible on both the CT slice and the photograph, such as the image pair displayed in Figure \ref{fig:photo_ct_comparison}. The apple core can sometimes be used as a shared visual feature because it is visible in CT and photographs but not in all slices. Thirdly, the alignment error is unknown. For small features (e.g. sepal or petal bundles), the alignment error may be larger than the feature size. In such cases, finding the corresponding position on the CT slice is challenging, and there is a risk of finding a false correspondence. Knowing the alignment error gives an indication of which size of features can reliably be matched.

In this paper, we present a workflow for creating datasets of slice photographs and their corresponding CT slices, which solves the three issues mentioned above. Image registration (automatic alignment) was used instead of manual alignment, making it possible to create large datasets with little extra effort. The alignment quality is not affected if a slice has little shared visual information inside of the apple (Fig. \ref{fig:photo_ct_comparison}), because the image registration only uses the outer shape of the apple. Moreover, the alignment error can be calculated by using the position of the apple core, because the apple core was not used for the image registration. This paper is the first work where image registration is applied to 3D scans and photographs of sliced fruit. We expect that with slight modifications, a similar workflow could be applied to other fruits or MRI scans instead of CT scans.

There are two additional scientific contributions in this paper. Firstly, the problem formulation of the image registration method is new. It uses a task-specific transformation model to describe the alignment of multiple parallel 2D slices to a 3D volume, which is optimized jointly by using automatic differentiation. In this approach, information is shared between multiple slices, which improves the overall registration accuracy. Secondly, the presented workflow has been applied to create a dataset of 1347 slice photographs acquired from 107 ‘Kanzi’ apples with registered CT slices, which was made public for further research.

\section{Materials and methods}
\label{sec:materials_and_methods}
%Provide sufficient details to allow the work to be reproduced by an independent researcher. Methods that are already published should be summarized, and indicated by a reference. If quoting directly from a previously published method, use quotation marks and also cite the source. Any modifications to existing methods should also be described.

\subsection{Apple samples}
\label{subsec:kanzi_dataset}
In 2022, 120 ‘Kanzi’ apples that had been stored under CA conditions (4 °C, 1 kPa $\text{O}_\text{2}$, 1.5 kPa $\text{CO}_\text{2}$) for 8-9 months were obtained from FruitMasters, The Netherlands. The fruit was grown in orchards surrounding Geldermalsen, the Netherlands, and harvested at physiological maturity in 2021.

\subsection{Workflow overview}
Image registration methods use shared information between images to find a transformation function $T$ that describes the relative alignment between the images. In this paper, the image registration problem is finding a transformation function from the slice photographs to the CT scan in order to sample the corresponding CT slice for each photograph. CT scans and photographs have relatively little shared information (Fig. \ref{fig:photo_ct_comparison}), but the outer shape of the apple is visible on both CT and photographic images. Therefore, the image registration method was designed to use the outer shape to find the transformation function. However, by only using the outer shape, relatively little information is available on each slice photograph. To overcome this, the image registration method optimizes the alignment of all slices jointly. Apples can be sliced into parallel slices, causing all slices to be rotated equally. Therefore, only one set of rotation parameters is sufficient to describe the rotation of all slices relative to the CT scan. The image registration method optimizes these parameters so that the outer shape matches well with the CT scan on all slices. This means that the combined information from all slices is used to find the rotation parameters and other parameters that are shared between all slices.
\begin{figure}[htb]
    \centering
    \includegraphics[width=0.95\textwidth]{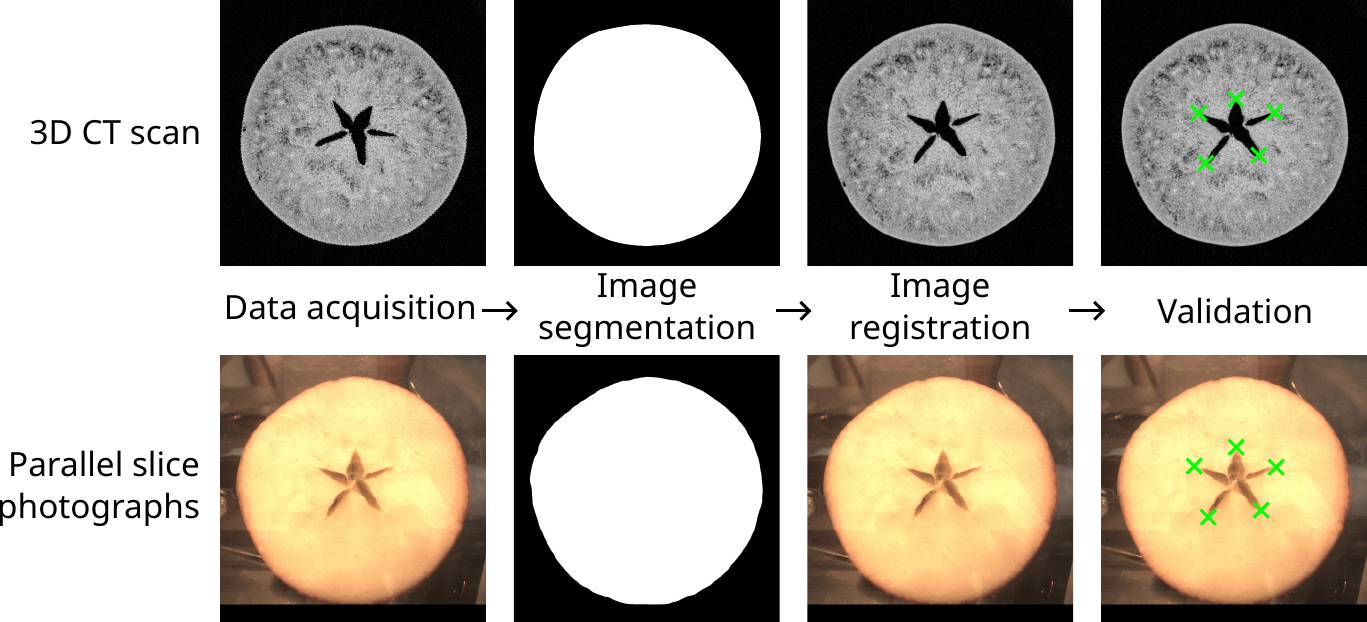}
    \caption{Overview of the workflow. For every step, one CT slice and photograph are shown to illustrate the results from that step. Note: In the actual workflow the CT scan is 3D data and there are multiple slice photographs.}
    \label{fig:workflow_overview}
\end{figure}

The complete workflow to obtain registered image pairs of apples consists of four steps: Data acquisition, image segmentation, image registration, and validation (Figure \ref{fig:workflow_overview}). In the data acquisition step, the fruit is first CT scanned and then sliced into parallel slices and photographed. In the image segmentation step, the outer shape of the apple is determined in both the CT scan and the slice photographs. In the image registration step, the transformation function from the slice photographs to the CT scan is optimized based on the outer shape of the apple. In the validation step, the registration error is measured from annotations of the apple's core. All steps are implemented as code in the Python programming language.

\subsection{Step 1: Data acquisition}
\label{subsec:data_acquisition}
The goal of this step was to acquire CT scans and slice photographs of the apple. We first CT scanned the apples. After that, we used a setup to slice and photograph the apple. This setup ensured that the slices were parallel and equally spaced and that the apple did not rotate during slicing.

\subsubsection{X-ray computed tomography(CT) acquisition and reconstruction}
\label{subsec:ct_acquisition}
The apples were scanned at the FleX-ray laboratory \citep{coban2020explorative} using a custom scanner developed by TESCAN-XRE, Gent, Belgium. A cone beam geometry with a circular trajectory was used to acquire 1440 projection images at an exposure time of 100ms, a tube peak voltage of 90\unit{\kilo\volt}, and a current of 550\unit{\micro\ampere}. Volumes were reconstructed with the FDK algorithm \citep{feldkamp1984practical}. To limit dataset size and scan time, 2 times detector pixel binning was used, resulting in a voxel size of 129.3\unit{\micro\meter}. Beam hardening correction was used from the FleXbox package \citep{kostenko2020prototyping}. All apples were scanned with the stem side on top. Moreover, a line was drawn on all apples from the stem to the calyx. The apples were put in the CT scanner so that the line was facing the X-ray source.

\subsubsection{Slicing and photograph acquisition}
\label{subsec:photo_acquisition}
\begin{figure}[htb]
    \centering
    \includegraphics[width=0.6\textwidth]{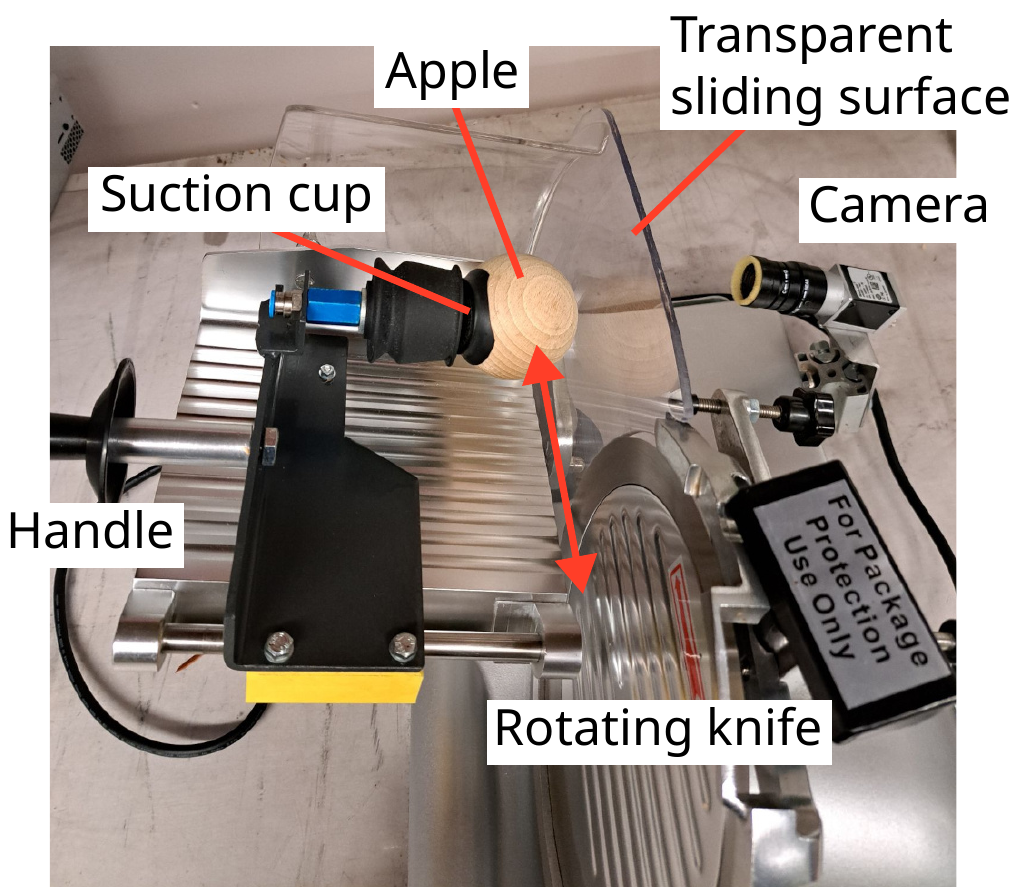}
    \caption{Setup used for slicing and photographing the apples.}
    \label{fig:slicing_machine}
\end{figure}

One day after CT scanning, the apples were sliced using a modified meat-slicing machine (CaterChef, EMGA, Mijdrecht, The Netherlands) (Figure \ref{fig:slicing_machine}). The sliding surface of the meat-slicing machine was replaced by a transparent acrylic sheet. A camera was placed behind the sliding surface, facing the apple. While in the machine, each apple was kept in place by a suction cup so it could not rotate during the slicing. All apples were sliced from the stem end to the calyx end with a slice thickness of approximately 4 \unit{\milli\meter}. Every time before slicing, a picture was taken of the remaining part of the apple through the transparent sliding surface. To ensure that the slice photographs were already roughly aligned with the CT scans, the apples were oriented in the slicing machine so that the line drawn earlier was on top. After each apple was sliced, the slices were visually inspected for brown tissue. At a later time, the type of browning was also noted for each fruit based on the photographs.

Sometimes, parts of the apple would break off while slicing the last few slices. The corresponding photographs were excluded. For three apples, the cutting damage already started halfway through the apple, and these apples were therefore excluded completely. Two apples were excluded because they had severe decay. The first eight apples were used to experiment with the lighting of the slicing setup. Because of the inconsistent lighting of the photographs of these apples, they were also excluded. After the exclusions, 107 apples remained.

\subsection{Step 2: Image segmentation}
\label{subsec:image_segmentation}
The goal of this step was to acquire a segmentation mask for each photograph and the CT scan. A segmentation mask is a black-and-white image of the same size as the original image, where a value of 1 indicates this part of the image shows the inside of the apple, and a value of 0 indicates this part of the image shows the outside of the apple. The segmentation masks were used in the image registration step to represent the outer shape of the apple.

\subsubsection{CT segmentation}
\label{subsec:ct_segmentation}
Thresholding was used to segment the shape of the apple from the CT scan. The threshold was determined by applying Otsu's method \citep{otsu1979threshold} to all CT scans individually and taking the average of the results. This average Otsu threshold was applied to all CT scans. Morphological closing, selection of the largest connected component, and hole filling were used as post-processing.

\subsubsection{Deep learning based photo segmentation}
\label{subsec:photo_segmentation}
\begin{figure}[htb]
    \centering
    \includegraphics[width=0.6\textwidth]{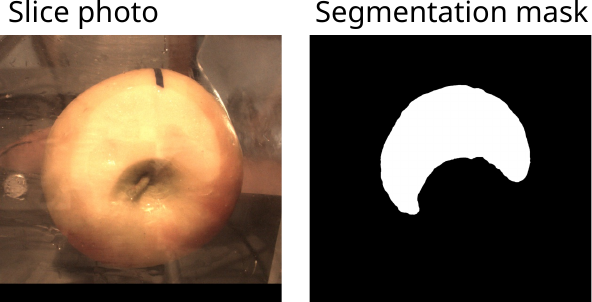}
    \caption{(Left) Slice photograph showing both the inside and the outside of the apple. (Right) Segmentation mask from the training set indicating the inside of the apple.}
    \label{fig:peel_on_photo}
\end{figure}

Segmenting the apple photographs could not be achieved with a single threshold. Only the outer shape of the apple should be segmented, which in the pictures is the part where the apple was sliced. However, the colors of the inside of the apple were very similar to the background or the outside of the apple in some photographs (Figure \ref{fig:peel_on_photo}). Deep learning has shown excellent results on various segmentation tasks \citep{wang2022medical}, so it was used for segmentation.

The network architecture was a Mixed-scale dense convolutional neural network (MSD-net) \citep{pelt2018mixed} implemented in PyTorch using the msd\_pytorch library \citep{hendriksen-2019-msd-pytor} using 200 layers. The normalized RGB values were used as the three input channels of the network. The network had one output channel to which a sigmoid activation function was applied. The binary cross entropy was used as a loss function, and the ADAM optimizer \citep{kingma2014adam} was used to minimize the loss function. For training the neural network, 175 photos were manually segmented. Data augmentation was applied to the training data using the \texttt{ColorJitter}, \texttt{ShiftScaleRotate} and \texttt{GaussNoise} functions of the Albumentations package \citep{info11020125}. Cross-validation was used to train five neural network instances using 80\% of the data as the training set and 20\% of the data as the validation set. The validation set was used for early stopping to reduce overfitting without compromising model accuracy. The outputs of the five neural network instances were converted into a segmentation mask by taking the average and applying a threshold of $0.5$. Selection of the largest connected component and hole filling were used as post-processing.

\subsection{Step 3: Image registration}
\label{subsec:image_registration}
The goal of this step was to find a transformation function $T$ that describes the relative alignment of the slice photographs to the CT scan. A task-specific transformation model was developed with shared parameters between all slices because the slices are parallel and equally spaced. The parameters of the transformation model were optimized based on the outer shape of the apple. This optimization was done in two steps: an initialization step and a full optimization step.

\subsubsection{Transformation model}
\label{subsec:transformation_model}
\begin{figure}[htb]
    \centering
    \includegraphics[width=0.65\textwidth]{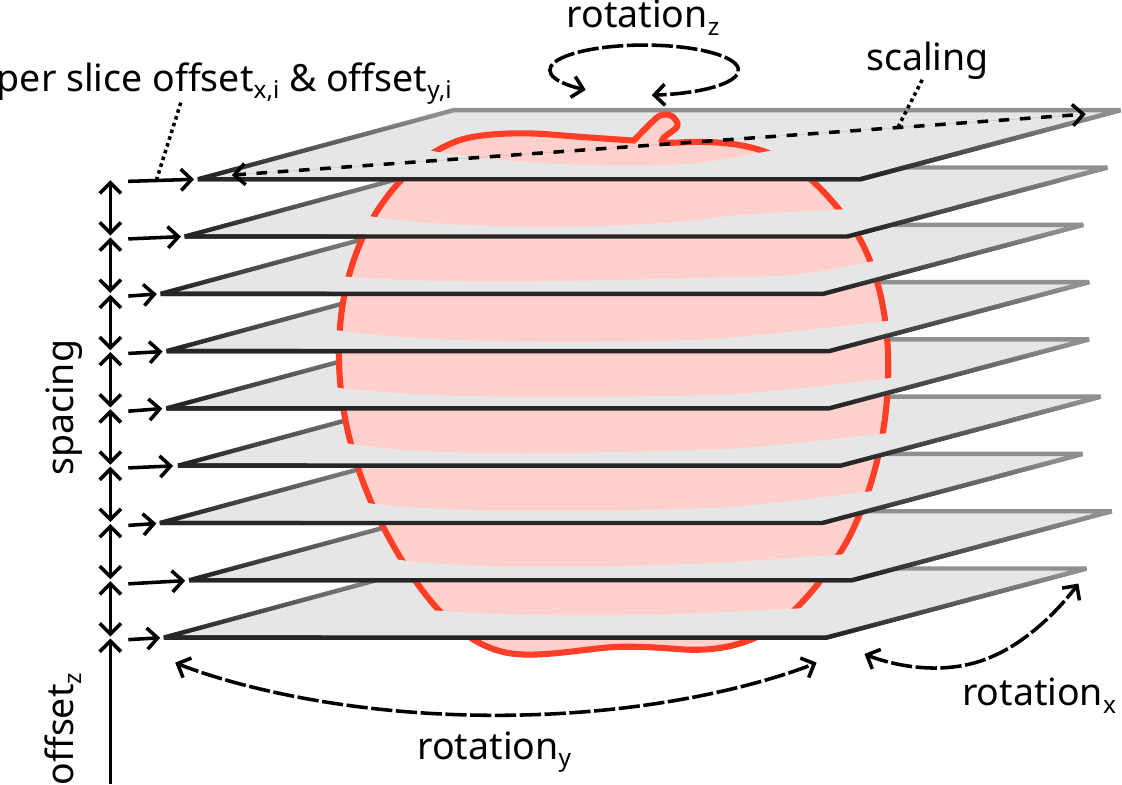}
    \caption{Illustration of the parameters of the parallel slice transformation model. Only ‘‘$\text{offset}_{x,i}$" and ‘‘$\text{offset}_{y,i}$" are unique for each slice. The other parameters are shared between all slices of the same apple.}
    \label{fig:transformation_model}
\end{figure}

Transformation functions can include different transformations such as rotating, shifting, scaling, and shearing. The transformation model defines which transformations are included and how they are parameterized. A transformation model was developed specifically for this image registration task. It describes how the parallel 2D slice photographs are aligned relative to each other and to the 3D CT scan. Most transformation parameters are shared between all photographs. These are: a scalar ‘‘$\text{scaling}$" to match the photograph pixel size to the CT voxel size; a scalar ‘‘$\text{spacing}$" specifying the $z$-axis distance between adjacent photographs; a scalar ‘‘$\text{offset}_z$" specifying the $z$-axis offset of the whole apple; and three scalars ‘‘$\text{rotation}_x$", ‘‘$\text{rotation}_y$", ‘‘$\text{rotation}_z$" describing the rotation as Euler angles. The non-shared parameters are two scalars ‘‘$\text{offset}_{x,i}$" and ‘‘$\text{offset}_{y,i}$" for each photograph specifying the offset along the $x$ and $y$-axes. The transformation model is illustrated in Figure \ref{fig:transformation_model}. The order of applying the transformations is as follows for a given coordinate vector $c$ in the $i$-th slice photograph:

\begin{equation}
    T(\bm{c}, \bm{\theta}, i) = \bm{R}_{\bm{\theta}} \left ( \left [
    \begin{array}{c}
        \text{offset}_{x,i} \\
        \text{offset}_{y,i} \\
        \text{offset}_z + \text{spacing} \cdot i
    \end{array}
    \right ] + \text{scaling} \cdot \boldsymbol{c} \right ).
\end{equation}

All transformation parameters are contained in vector $\bm{\theta} = $ \{ $\text{rotation}_x$, $\text{rotation}_y$, $\text{rotation}_z$, $\text{scaling}$, $\text{spacing}$, $\text{offset}_{x,1}$, $\text{offset}_{y,1}$, $\text{offset}_{x,2}$, $\text{offset}_{y,2}$, $...$ \}. $\bm{R}_{\bm{\theta}}$ is the rotation matrix based on the $\text{rotation}_x$, $\text{rotation}_y$ and $\text{rotation}_z$ parameters in $\bm{\theta}$.

\subsubsection{Initialization of the transformation parameters}
\label{subsec:initialization}
Before performing the full optimization of the transformation parameters, the parameters are initialized. Initializing the parameters avoids converging to the wrong local optimum and saves some computation time.

The rotation parameters were initialized to zero. Because of the way the position of the stem and the drawn line were used during acquisition (Sections \ref{subsec:ct_acquisition} \& \ref{subsec:photo_acquisition}) the rotation was already roughly aligned. 

The ‘‘$\text{scaling}$", ‘‘$\text{offset}_z$", and ‘‘$\text{spacing}$" parameters were initialized by matching the area of the segmentations in horizontal slices between the CT and the photographs. We call this the area profile. The area profile of the photographs was defined as the set of points $\boldsymbol{p}_i$ with one point for each slice photograph, where $i$ stands for the index of the photograph and $A_{photo}(i)$ for the segmentation area of the photograph at index $i$:
\begin{equation}
    \boldsymbol{p}_i = \left [
    \begin{array}{c}
        \sqrt{A_{\text{photo}}(i)}\cdot\text{scaling} \\
        \text{offset}_z + \text{spacing} \cdot i
    \end{array}
    \right ].
\end{equation}

The area profile of the CT scan was defined as a set of points $\boldsymbol{s}_j$ with one point for each horizontal slice of the CT scan, where $j$ stands for the index of the horizontal CT slice and $A_{CT}(j)$ for the segmentation area of slice $j$.
\begin{equation}
    \boldsymbol{s}_j = \left [
    \begin{array}{c}
        \sqrt{A_{\text{CT}}(j)} \\
        j
    \end{array}
    \right ].
\end{equation}
The mean over the squared distances from every point in the photo area profile to the closest point in the CT area profile was used as a cost function. This cost function was optimized using gradient descent with momentum \citep{rumelhart1985learning, goh2017why}, and the gradient was automatically derived using PyTorch \citep{paszke2017automatic}. This approach is similar to the SGD-ICP method \citep{maken2019speeding}. The learning rates were manually tuned for each parameter to be as large as possible without causing exploding gradients. A momentum of 0.6 was used on all parameters. A fixed number of 10000 iterations was used.

The ‘‘$\text{offset}_{x,i}$" and ‘‘$\text{offset}_{y,i}$" parameters were initialized so that the center of mass of each slice photograph segmentation was in the same position as the center of mass of the CT segmentation slice that was sampled based on the current ‘‘$\text{scaling}$", ‘‘$\text{offset}_z$" and ‘‘$\text{spacing}$" estimates.

\subsubsection{Joint optimization of the transformation parameters}
\label{subsec:optimization}
After initializing the transformation parameters, a more precise optimization was performed to get to the final image registration result. This step uses the mean square error(MSE) between the segmentations of the CT scans and slice photographs as a cost function. To evaluate the cost function, the positions of all pixels in the photograph segmentation masks were transformed into CT space using the current transformation parameters, and lookups were done in the CT segmentation mask using trilinear interpolation.

The cost function was optimized using gradient descent with momentum \citep{rumelhart1985learning, goh2017why}. The gradient was automatically derived using PyTorch \citep{paszke2017automatic}, and all calculations were performed on the GPU. The learning rates were manually tuned for each parameter to be as large as possible without causing exploding gradients, and a momentum of 0.75 was used on all parameters. The stopping criterion was that over the last 1000 iterations, the minimum and maximum values of the cost function were less than 0.00001 apart.

\subsection{Step 4: Validation}
\label{subsec:validation}
The goal of this step was to measure the registration error. The segmentation error was also measured because the registration was optimized based on segmentation masks.

\subsubsection{In plane core endpoint distance (IPCED)}
\label{subsec:IPCED}
\begin{figure}[htb]
    \centering
    \includegraphics[width=0.8\textwidth]{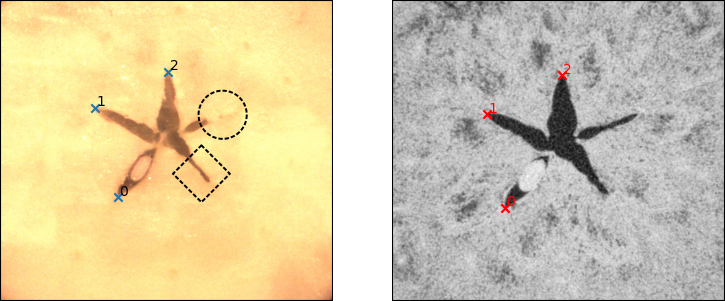}
    \caption{Example of the annotations of the core endpoints used to measure the registration error between a slice photograph (left) and the registered CT slice (right). When the endpoint of a part of the core was unclear (circular region) or when the core structure was very thin (diamond region) the endpoints were not annotated.}
    \label{fig:core_annotations}
\end{figure}

The core of the apple is visible on both the CT scans and the slice photographs and was not used for optimizing the registration parameters. Therefore, the core could be used to measure the registration error. The outer corner points of the apple core were selected manually on the photographs. Later, the closest corresponding points on the registered CT slice were manually selected (Figure \ref{fig:core_annotations}). The in-plane core endpoint distance (IPCED) was defined as the average distance between the annotations in the slice photograph and the registered CT slice.

To minimize the effect of the out-of-plane registration error, the points were only selected on one slice photograph close to the center of the apple core because, at its center, the edges of the core are roughly perpendicular to the slicing plane. Moreover, very thin core structures or those with an unclear endpoint were not annotated. To avoid bias, the photographs were selected and the annotations on the photographs were done without looking at the CT scans. The CT point annotations were placed with the annotated photograph and the registered CT scan side-by-side in a view similar to Figure \ref{fig:core_annotations}.

\subsubsection{Segmentation error}
\label{subsec:segmentation_quality}
The segmentation error was evaluated to understand how much of the registration error was caused by the quality of the segmentation of the photographs. For this goal, a test set was created of 20 manually segmented slice photographs of 20 different apples that had not been used for training the segmentation neural network. The five neural network instances that were trained using different cross-validation splits were applied to these photographs. The postprocessing was applied to the mean of the five networks like in Section \ref{subsec:photo_segmentation} and to each network output separately. True positives ($TP$) represent pixels correctly segmented as being inside the apple; true negatives ($TN$) represent pixels correctly segmented as being outside the apple; false positives ($FP$) represent pixels incorrectly segmented as inside the apple; and false negatives ($FN$) represent pixels incorrectly segmented as being outside the apple. The following error metrics were calculated: accuracy $(TP + TN) \:/\: (TP + TN + FP + FN)$, precision $TP \:/\: (TP + TN)$, recall $TP \:/\: (TP + FN)$, and distance from the edge pixels in the neural network segmentation to the closest edge pixels in the manual segmentation.

\subsection{Validation of registration method}
\label{subsec:registration_experiments}
The registration method (Section \ref{subsec:image_registration}) has not been used before; thus, two additional experiments were performed to measure the added value of this new method.

\subsubsection{Number of slices experiment}
\label{subsec:subsets}
The slicing method (Section \ref{subsec:photo_acquisition}) was supposed to result in parallel slices, but slight deviations may occur that could accumulate when many slices are used. However, using many slices may also improve results by including more information in the optimization. The goal of this experiment was to explore how the number of slices used in the joint registration method related to the registration error. To vary the number of slices, not all available slices were used.

For every apple, slices at the top and the bottom of the apple were excluded to make subsets with a specific number of slices. To be able to measure the IPCED, the annotated center slice was always included. Additionally, the three slices above and two slices below the annotated slice were included one by one, creating a new subset with every included slice, resulting in six subsets for each apple. For 30 Apples, these slices were not all available, so those fruit were excluded from this experiment. On the subsets with fewer slices on some apples, the registration error was too large to do the annotations necessary for calculating the IPCED. To allow comparison between subset sizes these fruit were excluded from calculating the average IPCED on all subset sizes.

\subsubsection{Separate slice registration intersection experiment}
\label{subsec:intersection_experiment}
Because the previous experiment (Section \ref{subsec:subsets}) used the IPCED, it only calculated the error at one central slice. This experiment was performed to get insight into the registration robustness over the whole apple. Each apple's photographs were acquired from slices from top to bottom. Therefore, it is physically impossible that slices intersected each other, or that the position of later photographs was before the position of earlier photographs. Our joint registration method enforces that the slices are parallel and in the correct order, but in this experiment, each slice was registered separately to test what would occur if these properties were not enforced.

The same method was used as for joint registration, with a minor modification. In the transformation model, the $\text{offset}_z$ and $\text{spacing}$ parameters would offer the same degree of freedom when just one slice is used, so the $spacing$ parameter was kept constant. The initialization method from Section \ref{subsec:initialization}, and the full joint optimization method from Section \ref{subsec:optimization} were used to initialize the transformation model parameters.

\begin{figure}[htb]
    \centering
    \includegraphics[width=0.8\textwidth]{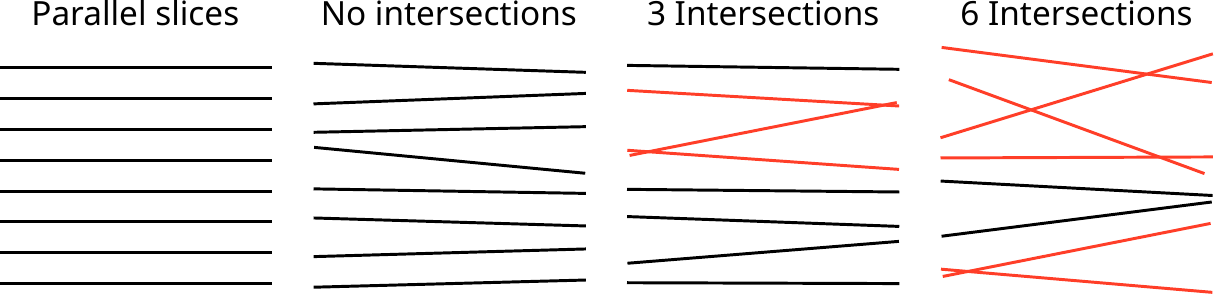}
    \caption{Illustration of the intersection test. The red lines indicate intersecting slices.}
    \label{fig:intersection_illustration}
\end{figure}

An intersection test was developed to test if the slices were parallel and in the correct order. First, the coordinates of every pixel inside each photo slice segmentation mask were transformed into CT space. A pixel belonging to slice $i$ should not be inside either of the two convex hulls formed by slices with indices larger and smaller than $i$. If any pixel in a slice was inside either of these convex hulls, that entire slice was marked as intersecting. A distinction was made between apples where three or fewer adjacent slices were intersecting and where more slices were intersecting. Figure \ref{fig:intersection_illustration} illustrates the intersection test.

\section{Results}
\label{sec:results}
%Results should be clear and concise statements of the results as documented by the data collected and their subsequent analysis.

\subsection{Registration and segmentation errors}
\label{subsec:kanzi_results}
The full workflow was applied to obtain the slice photographs and registered CT slices in the ‘Kanzi' data set. The average IPCED for all 107 apples was $1.47 \pm 0.40\:\unit{\milli\meter}$.

The calculated segmentation metrics are displayed in Table \ref{tab:segmentation_metrics}. The \textit{combined} row represents the segmentations that were used for registration. The average segmentation edge distance was $0.195 \pm 0.104\:\unit{\milli\meter}$.

\begin{table}[htb]
\centering
\caption{Segmentation quality metrics of the neural network instances trained on different training and validation data splits.} 
\label{tab:segmentation_metrics}
\begin{tabular}{r|l|l|l|l}
 & \textbf{Accuracy (\%)} & \textbf{Precision (\%)} & \textbf{Recall (\%)\hspace{0.3cm}} & \textbf{Edge distance (mm)} \\ \hline
instance 0 & $99.37 \pm 0.30$ & $99.05 \pm 1.07$ & $99.41 \pm 0.63$ & $0.234 \pm 0.139$ \\ \hline
instance 1 & $99.26 \pm 0.50$ & $99.77 \pm 0.27$ & $98.45 \pm 1.57$ & $0.259 \pm 0.180$ \\ \hline
instance 2 & $99.37 \pm 0.26$ & $99.10 \pm 0.88$ & $99.39 \pm 0.65$ & $0.241 \pm 0.133$ \\ \hline
instance 3 & $99.33 \pm 0.44$ & $99.61 \pm 0.36$ & $98.78 \pm 1.41$ & $0.247 \pm 0.206$ \\ \hline
instance 4 & $99.39 \pm 0.33$ & $99.52 \pm 0.47$ & $99.04 \pm 1.05$ & $0.222 \pm 0.134$ \\ \hline
combined & $99.43 \pm 0.30$ & $99.54 \pm 0.42$ & $99.12 \pm 0.94$ & $0.195 \pm 0.104$ \\ \hline
\end{tabular}
\end{table}

\subsection{Validation of registration method}
\label{subsec:joint_results}
The results of the number of slices experiment are displayed in Table \ref{tab:subset_results}. The results show that the average IPCED decreased with every increase in the subset size. Moreover, only when a subset size of three or lower was used there were cases where the registration error was too large to do the annotations necessary for calculating the IPCED.

\begin{table}[htb]
    \centering
    \caption{Registration results when using different numbers of slices in the joint registration.}
    \label{tab:subset_results}
    \begin{tabular}{r|l|l}
    \begin{tabular}[c]{@{}r@{}}\textbf{No. of slices}\\\textbf{in subset}\end{tabular} & \begin{tabular}[c]{@{}r@{}}\textbf{No. of fruit where IPCED} \\ \textbf{could not be calculated} \end{tabular} & \textbf{IPCED (mm)} \\ \hline
    1 & 3 & $ 1.97 \pm 0.89$ \\ \hline
    2 & 7 & $ 1.83 \pm 0.80$ \\ \hline
    3 & 3 & $ 1.83 \pm 0.82$ \\ \hline
    4 & 0 & $ 1.75 \pm 0.81$ \\ \hline
    5 & 0 & $ 1.74 \pm 0.83$ \\ \hline
    6 & 0 & $ 1.71 \pm 0.86$ \\ \hline
    All & 0 & $ 1.52 \pm 0.39$ \\ \hline
    \end{tabular}
\end{table}

The results of the intersection experiment are displayed in Table \ref{tab:intersection_results}. The results show that intersections occur when registering the slices separately. Using the full joint registration method as an initialization instead of the normal profile initialization method decreased the number of intersections, but intersections remained in 32 of the apples.

\begin{table}[htb]
    \centering
    \caption{Results of the separate slice registration intersection experiment.}
    \label{tab:intersection_results}
    \begin{tabular}{r|l|l|l}
    \textbf{Initialization} & \textbf{0 Intersections} & \textbf{$\bm{\leq 3}$ Adjacent intersections} & \textbf{More intersections} \\ \hline
    Profile (Sec. \ref{subsec:initialization}) & 42 (39.2\%) & 42 (39.2\%) & 23 (21.5\%) \\ \hline
    Joint (Sec. \ref{subsec:optimization}) & 75 (70.0\%) & 24 (22.4\%) & 8 (7.5\%) \\ \hline
    \end{tabular}
\end{table}

\begin{figure}
    \centering
    \includegraphics[width=0.75\textwidth]{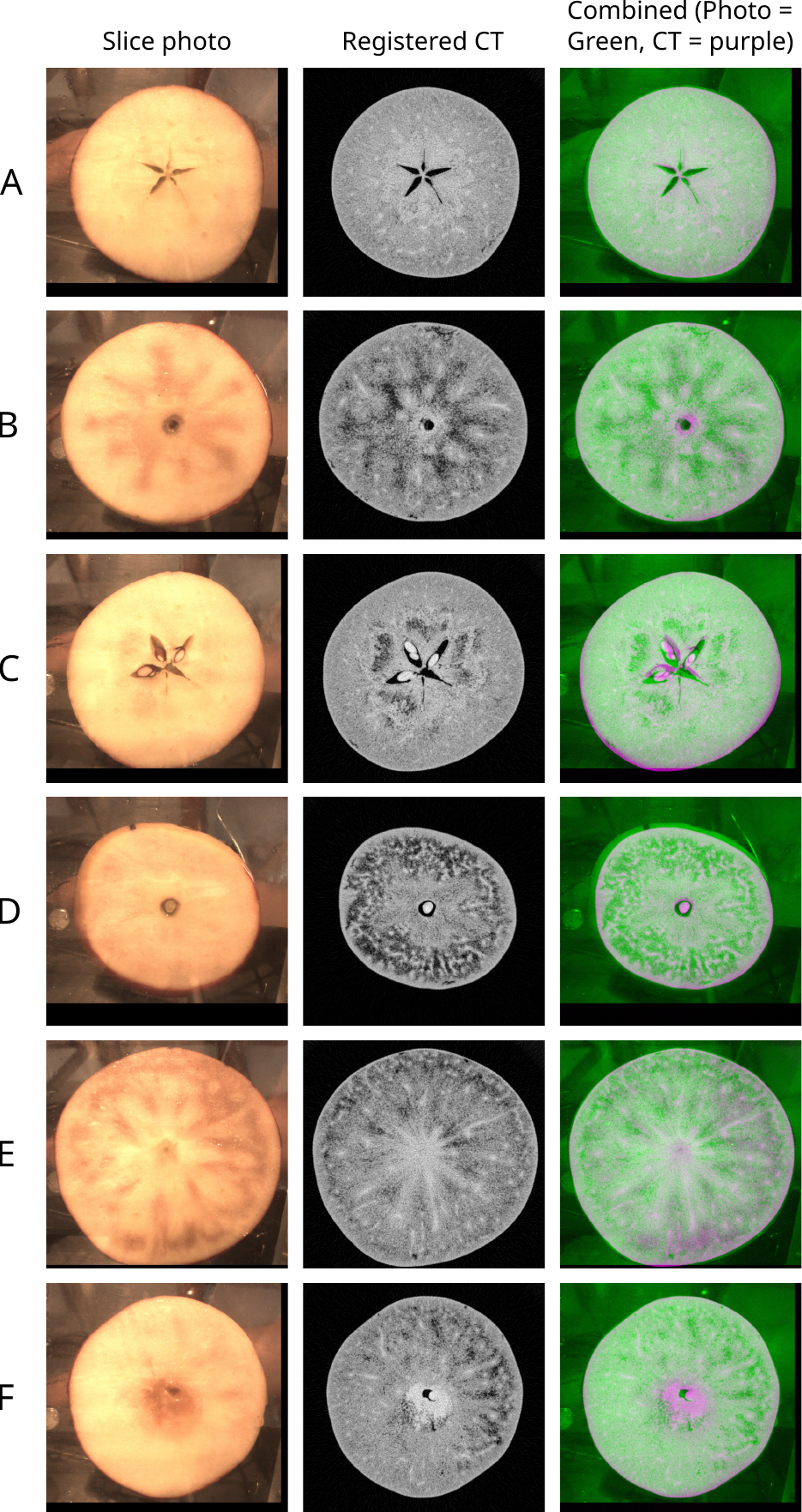}
    \caption{Examples images of a healthy apple and apples with different types of disorders. A: healthy, B: radial browning, C: core browning, D: senesence browning, E: diffuse flesh browning, F: decay. The combined view shows both images in the same plot and can be used to see the distance between edges. The combined view is green when the photo is brighter, purple when the CT slice is brighter, and a shade of grey when both images have similar intensities.}
    \label{fig:disorder_example}
\end{figure}

\begin{figure}
    \centering
    \includegraphics[width=0.75\textwidth]{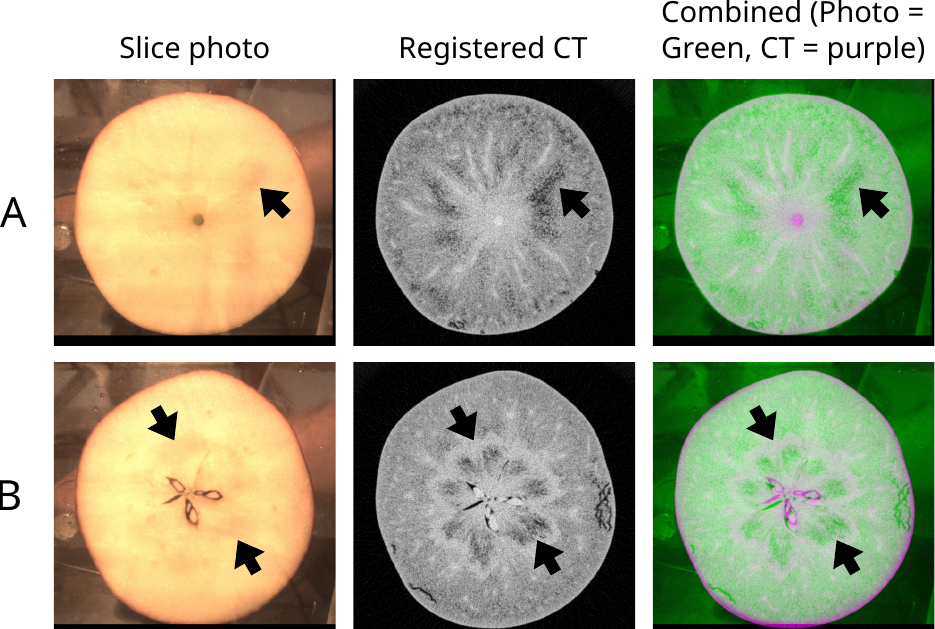}
    \caption{Examples of ‘Kanzi’ apples that were classified as healthy after cutting, but contained regions of flesh that appeared brown in CT images. The arrows indicate regions of suspected brown tissue, which were confirmed using the registered CT slice. The combined view shows both images in the same plot and can be used to see the distance between edges. The combined view is green when the photo is brighter, purple when the CT slice is brighter, and a shade of grey when both images have similar intensities.}
    \label{fig:example_arrows}
\end{figure}

\subsection{Disorder evaluation}
\label{subsec:browning_appearance}
Of the 107 ‘Kanzi’ apples in the dataset, 45 were healthy after storage, and 62 contained brown regions of flesh with varying levels of severity. Senescent breakdown, radial flesh, diffuse flesh, and core browning were all identified. An example of each browning type is presented in Figure \ref{fig:disorder_example}. We found that dark regions in CT scans corresponded to regions of brown flesh in ‘Kanzi’ fruit. High-intensity regions occurred on the border of dark regions in CT scans for fruit with core browning (Figure \ref{fig:disorder_example}C). Fruit with decay displayed areas of high and low-intensity regions in CT scans (Figure \ref{fig:disorder_example}F). The wet, decayed flesh had a high voxel intensity, and the remaining flesh resembled senescence browning (Figure \ref{fig:disorder_example}D).

During the browning classification, several fruit marked as healthy during slicing displayed slightly darker areas in the photo images. However, it was difficult to determine with the human eye whether these areas were brown tissue or shadows. In six apples a slightly darker area in the photo images coincided with a dark region in the CT scan (Figure \ref{fig:example_arrows}). Based on the CT scan and photo together, we expect that these apples had mild browning.

\subsection{Data and code}
The CT scans, slice photographs, and browning status are publicly available on Zenodo \citep{dirk_elias_schut_2023_8167285}. All data used in the results section, including registration and segmentation results, are publicly available as a separate dataset \citep{zenodo_results}. The Python code is available on Github \citep{workflow_github}.

\section{Discussion}
\label{sec:discussion}
%This section should confront the results of the work to the state of the art, explore their significance but not contain a restatement of them. It is advised to not combine the Results and Discussion section. Avoid extensive citations and discussion of published literature.

\subsection{Physiological disorders}
\label{subsec:kanzi_discussion}
Most physiological disorders have affected regions much larger than the average IPCED of 1.47 mm. Therefore, the provided photo-CT image pairs should be suitable for disorder evaluation. The photographs can be used as ground truth to understand the appearance of disorders in CT. Registered image pairs can also provide complementary information, which could be explored in future work. Combining information from registered CT, MRI, or PET scans has been used in medical imaging to diagnose diseases such as cancer \citep{blodgett2007pet, du2016overview}.

Our finding that dark regions in CT scans corresponded to regions of brown flesh is consistent with previous reports \citep{HERREMANS2013114, VANDAEL2019218}. Additionally, high-intensity regions surrounding the dark regions in fruit with core browning, were also found by \citet{lammertyn2003mri} study on pears. The dark regions in affected fruit are due to the structural collapse of cells and the moment of water away from affected flesh, reducing voxel intensity \citep{HERREMANS2013114, VANDELOOVERBOSCH2020107170}. \citet{HERREMANS2013114} suggest that the available water will diffuse towards the fruit peel, where it is lost to the environment. Water-rich tissue has a high voxel intensity \citep{DIELS201724}; thus, the high-intensity regions surrounding the dark regions in fruit with core browning are likely caused by the movement of water away from the affected flesh and towards the fruit’s surface. On six apples, browning was detected on the CT scan but not during visual inspection, which suggests that CT can potentially be used for earlier or more reliable detection of browning than visual inspection. Early non-destructive detection of browning would be commercially valuable. \cite{part2} demonstrated that the level of browning in an apple can be determined from the percentage of dark spots in a single CT slice.

The high-intensity regions of decayed flesh in our CT scans are similar to those obtained from fruit with watercore \citep{HERREMANS2013114}. Watercore is a physiological disorder associated with the flooding of air-filled pores with aqueous fluid, increasing voxel intensity in CT scans \citep{HERREMANS2013114}. \citet{HERREMANS2013114} study successfully identified the water-soaked regions typical for fruit affected with watercore using CT with up to 89 \% accuracy. As watercore and decay appear similar under CT, fruit with decayed flesh may be successfully identified using CT and image thresholding. However, distinguishing between regions of flesh suffering from decay and watercore might be challenging due to their similar appearance. Although, the separation of decay and watercore may only be an issue in growing regions where apples with watercore are a desired quality trait and are sold at a premium.

\subsection{Data acquisition and image segmentation}
After using the workflow on the ‘Kanzi’ dataset, several areas of improvement have been identified for acquiring sets of slice photographs to make image registration to CT or MRI scans simpler and more accurate.

Firstly, the scale parameter could be derived directly from the image instead of being fitted in the registration step. For example, a ruler could be glued to the acrylic sliding surface of the slicing machine. Furthermore, if the direction of the camera is not perpendicular to the sliding surface, the scale will vary over the image because of the perspective projection. Perspective distortion can be corrected by placing a square frame as a marker around the area where the apple will be photographed \citep{jagannathan2005perspective}.

Secondly, the photo segmentation could be simplified if the contrast could be improved between the inside of the apple and both the peel and the background. If the contrast can be increased sufficiently, classical methods for segmentation might be used instead of a neural network. This would free up the time required for manually labeling images and save the costs for hardware and electricity of neural network training. Moreover, more accurate segmentation masks may be achieved. It may even be possible to remove the segmentation step completely by using a multi-modality cost function such as mutual information during the image registration \citep{maes1997multimodality}.

Improving contrast with the background can easily be achieved by placing an evenly colored surface of a contrasting color (e.g. blue or black) behind the machine. One suggestion for improving the contrast with the peel would be to paint the apple in a contrasting color before cutting. Another suggestion would be to cut the apple through the middle at the widest point and then put the apple in the machine in two parts with the cutting plane facing the camera. This way the apple would get narrower further away from the camera so that the peel would not be visible in the photographs. The middle slice would be photographed twice, and by registration between these two images, the rotation difference between the sets of photographs of the two halves of the apple could be compensated.

\subsection{Image registration}
The idea of segmentation-based registration has been applied before by \citet{museyko2010binary}, which showed that the results for image registration between two CT scans of the human femur or lumbar spine were improved by segmenting the bone structures before doing image registration. Segmentations were also used to overcome the differences in appearance while registering a 3D ultrasound scan and a CT scan of the liver \citep{schut2018automatic}. Pears have a more irregular shape than apples, so performing segmentation-based registration on pears may be more accurate or require less data.

Task-specific transformation models have been used in medical image registration of the heart \citep{heyde2013three} and liver \citep{wein2008automatic} to allow for specific motions with less risk of overfitting than using a more general transformation model. The most common approach for image registration is to use a toolkit such as Elastix \citep{klein2009elastix} or ITK \citep{avants2014insight}, where users can combine high-quality implementations of commonly used transformation models, cost functions, and optimizers. Many optimizers use the gradient, so gradient computation code has to be provided for each cost function and transformation model. While toolkits make it easy to prototype combinations of existing methods, the interactions with existing C++ code and the necessity of manually writing gradient code introduce a high barrier to adding custom transformation models or cost functions. The use of PyTorch for automatic differentiation \citep{paszke2017automatic} in this work lowered this barrier considerably. AirLab \citep{sandkuhler2018airlab} is a recently developed image registration toolkit that uses PyTorch automatic differentiation internally. In future work on custom transformation models, the existing AirLab implementations of optimizers and cost functions could be used.

Combining information from multiple slices has been done in medical histology image registration \citep{PichatIYOM18, FerranteP17}. In medical histology, a tissue sample is sliced into thin sections, which are put on glass slides, stained, and photographed under a microscope. Like apples in this paper, histological samples are sometimes scanned in a CT or MRI scanner before being sliced. Registration between histological slides and CT or MRI scans is therefore similar to the apple registration problem described in this paper. However, there are also differences. Large full-object deformations may be present in registration between histology and CT or MRI, so a deformable transformation model is used. To fit the many parameters of a deformable model, combining information from multiple histological photographs can be beneficial \citep{ceritoglu2010large}. The processes of slicing, mounting the slice on a glass slide, and staining can introduce artifacts and deformations in the slicing plane \citep{PichatIYOM18}. This makes it challenging to find the accurate relative positions of adjacent slices, which is called 3D reconstruction. Joint optimization of the 3D reconstruction and image registration problems can compensate for these effects. Joint optimization has been done by alternating between registering the slices individually to the CT or MRI scan to improve the 3D reconstruction and registering the whole reconstructed volume to the CT or MRI scan to improve the whole object registration \citep{malandain2004fusion, yang2012mri, goubran2013image, adler2014histology} and Section 5.1.4 in \citet{PichatIYOM18}. The method in this paper differs from those earlier methods in that the 3D reconstruction and registration problems are optimized within the same optimization problem.

\subsection{Validation}
The IPCED metric defined in Section \ref{subsec:IPCED} is easy to calculate for any apple or pear with only a few manual annotations. However, it is limited in that it only considers the in-plane error and can only be calculated at the core of the apple. For medical histology applications, needle tracks have been used as a marker for evaluation \citep{PichatIYOM18}. Inserting and removing multiple needles before data acquisition introduces straight holes in the apple. These are likely easy to segment in the CT scan and easy to detect in the photographs. The registration error can be calculated as the distance from the holes in the photograph to the nearest hole in the CT scan. Moreover, these features could be used directly for image registration. However, this metric would require damaging the apples and would introduce an extra sample preparation step and an extra detection step to the workflow.

\section{Conclusions}
\label{sec:conclusions}
%Authors are encouraged to succinctly state conclusions relative to the objectives of the study. The main conclusions may be presented in a short Conclusions section, which may stand alone or form a subsection within the Discussion section.

In this paper, we presented a workflow for creating datasets of registered image pairs of slice photographs and CT slices. Part of the workflow is a joint 2D-3D image registration method (Section \ref{subsec:image_registration}), which uses a task-specific transformation model to register all slices of one apple within a single optimization problem. It was shown that this joint approach is both more precise and more robust to large registration errors compared to image registration of separate slices (Section \ref{subsec:joint_results}). Using this workflow, a dataset was created of 1347 slice photographs acquired from 107 ‘Kanzi’ apples, which has been published alongside this paper. The image registration error of this dataset according to the in-plane core endpoint distance (IPCED) metric was $1.47 \pm 0.40\;\unit{\milli\meter}$ (Section \ref{subsec:kanzi_results}). The appearance of disorders in this dataset mostly matched the descriptions from earlier works on CT scanning of disordered apples (Section \ref{subsec:kanzi_discussion}). However, six apples that looked healthy directly after slicing showed dark regions on the CT scans associated with browning (Section \ref{subsec:browning_appearance}), indicating that CT may have the potential for detecting the early stages of browning.

\section*{Acknowledgements}
The authors would like to acknowledge Michiel Couvée for technical assistance with ‘Kanzi’ fruit slicing.  The authors also acknowledge TESCAN-XRE NV for their collaboration and support of the FleX-ray laboratory.

\section*{Funding}
This work was funded by the Dutch Research Council (NWO) through the UTOPIA project (ENWSS.2018.003).

\printbibliography

\end{document}